\documentclass{PoS}

\title{Search for Lepton Flavour Violation at HERA}

\ShortTitle{Search for Lepton Flavour Violation at HERA}

\author{\speaker{David M. South}\thanks{on behalf of the H1 Collaboration.}\\
  Deutsches Elektronen Synchrotron\\
  Notkestrasse 85\\
  22607 Hamburg, Germany\\
  E-mail: \email{david.south@desy.de}}

\abstract{
A search for second and third generation scalar and vector leptoquarks produced in
$ep$ collisions via the lepton flavour violating processes \mbox{$ep\rightarrow \mu  X$}
and \mbox{$ep\rightarrow \tau X$} is performed by the H1 Collaboration at HERA.
The full H1 $ep$ data sample taken at a centre-of-mass energy $\sqrt{s} = 319$~GeV
is used for the analysis, corresponding to an integrated luminosity of $411$~pb$^{-1}$.
No evidence for the production of such leptoquarks is observed in the H1 data.
Leptoquarks produced in $ep$ collisions with a coupling strength of
$\lambda=0.3$ and decaying with the same coupling strength to a muon-quark
pair or a tau-quark pair are excluded at $95\%$ confidence level up to
leptoquark masses of $712$~GeV and $479$~GeV, respectively.
}

\FullConference{36th International Conference on High Energy Physics,\\
		July 4-11, 2012\\
		Melbourne, Australia}

\begin{document}

%%%%%%%%%%%%%%%%%%%%%%%%%%%%%%%%%%%%%%%%%%%%%%%%%%%%%%%%%%%%
\section{Leptoquark production at HERA}
\label{sec:theory}

The $ep$ collisions at HERA provide a unique possibility to search for 
new particles coupling directly to a lepton and a quark.
Leptoquarks (LQs), colour triplet bosons that do just that, are an example
of such particles and appear in many theories attempting to unify the
quark and lepton sectors of the Standard Model (SM).

%%%

A discussion of the phenomenology of LQs at HERA can be found
elsewhere~\cite{Adloff:1999tp}.
In the framework of the Buchm\"uller-R\"uckl-Wyler (BRW) effective
model~\cite{BRW}, LQs are classified into $14$ types with respect to the
quantum numbers spin $J$, weak isospin $I$ and chirality $C$,
resulting in seven scalar ($J=0$) and seven vector ($J=1$) LQs.
Whereas all $14$ LQs couple to charged lepton-quark pairs, four of the left-handed LQs,
namely $S_0^L$, $S_{1}^L$, $V_0^L$ and $V_{1}^L$, may also decay to a
neutrino-quark pair.
In particular, for $S_0^L$ and $V_0^L$ the branching fraction of decays into a
charged lepton-quark pair is predicted by the model to be
\mbox{$\beta_\ell\!=\!\Gamma_{\rm \ell q}/(\Gamma_{\rm \ell q}+\Gamma_{\rm \nu_\ell q})\!=0.5$},
where \mbox{$\Gamma_{\rm \ell q}$} (\mbox{$\Gamma_{\rm \nu_\ell q}$}) denotes the partial
width for the LQ decay to a charged lepton (neutrino) and a quark $q$.
The branching fraction of decays into a neutrino-quark pair is then given by
$\beta_{\nu_{\ell}} =1-\beta_{\ell}$.

%%%

Leptoquarks carry both lepton ($L$) and baryon ($B$) quantum numbers, and the
fermion number \mbox{$F\!=\!L\!+\!3\,B$} is assumed to be conserved.
Leptoquark processes at HERA proceed directly via $s$-channel resonant LQ production
or indirectly via $u$-channel virtual LQ exchange.
A dimensionless parameter $\lambda$ defines the coupling at the
lepton-quark-LQ vertex.
For LQ masses well below the centre-of-mass energy $\sqrt{s}=319$~GeV,
the $s$-channel production of $F = 2$ ($F = 0$) LQs in $e^-p$ ($e^+p$) collisions dominates.
However, for LQ masses above $\sqrt{s}$, both the $s$ and $u$-channel processes are important
such that both $e^-p$ and $e^+p$ collisions have similar sensitivity to all LQs types.

%%%

The analysis presented here examines LQ decays to a quark and a second or third
generation lepton.
However, assuming a LQ produced at HERA observes flavour conservation, which is implicit in the
BRW model, then such a particle would decay exclusively into a quark and a first
generation lepton, $ep \rightarrow eX$ or $ep \rightarrow \nu_{e} X$.
Dedicated searches have been performed by H1 for such first generation LQs,
where the SM expectation is dominated by neutral current (NC) and charged current (CC) deep
inelastic scattering (DIS) background~\cite{Aktas:2005pr,h1lq2011}.
A search for first generation leptoquarks was recently performed using the complete
H1 data set, where for a coupling of electromagnetic strength
$\lambda = \sqrt{4 \pi \alpha_{\rm em}} = 0.3$, first generation LQs are excluded at $95\%$
confidence level (CL) up to leptoquark masses of $800$~GeV, depending on the leptoquark
type~\cite{h1lq2011,ichep2012lq}.

%%%%%%%%%%%%%%%%%%%%%%%%%%%%%%%%%%%%%%%%%%%%%%%%%%%%%%%%%%%%
\section{Search for lepton flavour violating leptoquarks}

A more general extension of the BRW model allows for the decay of LQs to final states
containing a quark and a lepton of a different flavour, i.e. a muon or tau lepton.
Non-zero couplings $\lambda_{eq_i}$ to an electron-quark pair and $\lambda_{\mu q_j}$
($\lambda_{\tau q_j}$) to a muon(tau)-quark pair are therefore assumed.
The indices $i$ and $j$ represent quark generation indices, such that $\lambda_{eq_i}$
denotes the coupling of an electron to a quark of generation $i$, and $\lambda_{\ell q_j}$
is the coupling  of the outgoing lepton (where $\ell = \mu$ or $\tau$) to a quark of
generation $j$.
Since neutrino flavours cannot be distinguished with the H1 experiment, final states
arising from the $ep \rightarrow \nu_{\mu} X$ or $ep \rightarrow \nu_{\tau} X$ processes
are not considered in the analysis.
An overview of this extended model for the LQ coupling to $u$ and $d$ quarks is
provided elsewhere~\cite{Aktas:2007ji}.

%%%

The introduction of lepton flavour violation (LFV) to leptoquark models would mean
that the processes $ep \rightarrow \mu X$ or $ep \rightarrow \tau X$, mediated by the
exchange of a second or third generation leptoquark would be observable at
HERA with final states containing a muon or the decay products of a tau lepton
back-to-back in the transverse plane with a hadronic system $X$.
The main SM background contribution to this topology is from photoproduction
events, in which a hadron is wrongly identified as a muon or a narrow hadronic jet
fakes the signature of the hadronic tau decay.
Similarly, the scattered electron in NC DIS events may also be misinterpreted as the
one-prong hadronic tau decay jet.
Smaller SM background contributions arise from events exhibiting intrinsic missing
transverse momentum (for example CC DIS), events containing high $P_{T}$ leptons
(such as lepton pair production, particularly inelastic muon-pair events if one muon
is unidentified) or events with both of these features (real $W$ production with
leptonic decay).
%

%%%

Searches for such signatures have been previously performed by
H1~\cite{Adloff:1999tp,Aktas:2007ji} and the latest analysis~\cite{h1lfv2011} is performed
using the complete $ep$ H1 collision data taken at a centre-of-mass energy
$\sqrt{s} = 319$~GeV, which was recorded during the years 1998-2007.
The corresponding integrated luminosity of $245$~pb$^{-1}$ for $e^{+}p$ collisions and
$166$~pb$^{-1}$ for $e^{-}p$ collisions represents an increase in size of the data sample
with respect to the previous publication by a factor of $3$ and $12$, respectively.

%%%

Leptoquarks with couplings to first and second generation leptons
may decay to a muon and a quark.
Therefore, event topologies with an isolated, high transverse momentum $P_{T}$ muon
back-to-back to a hadronic system in the transverse plane are selected.
Several additional cuts based on the transverse and longitudinal event balance are
employed to remove the SM background~\cite{h1lfv2011}.

%%%

Leptoquarks with couplings to first and third generation leptons may decay to a tau and a quark.
Tau leptons are identified using the muonic and one-prong hadronic decays of the tau.
The analysis of the muonic decay channel employs the same selection as in the second
generation LQ search.
The tau decay results in missing transverse momentum in the event due to the escaping
neutrinos and this is exploited in the analysis of the hadronic decay channel, which
also uses the track multiplicity of narrow jets to identify one-prong tau decay
jet candidates.
Full details of the event selection can be found in the H1 publication~\cite{h1lfv2011}.

%%%

\begin{figure}[t]
  \includegraphics[width=0.495\textwidth]{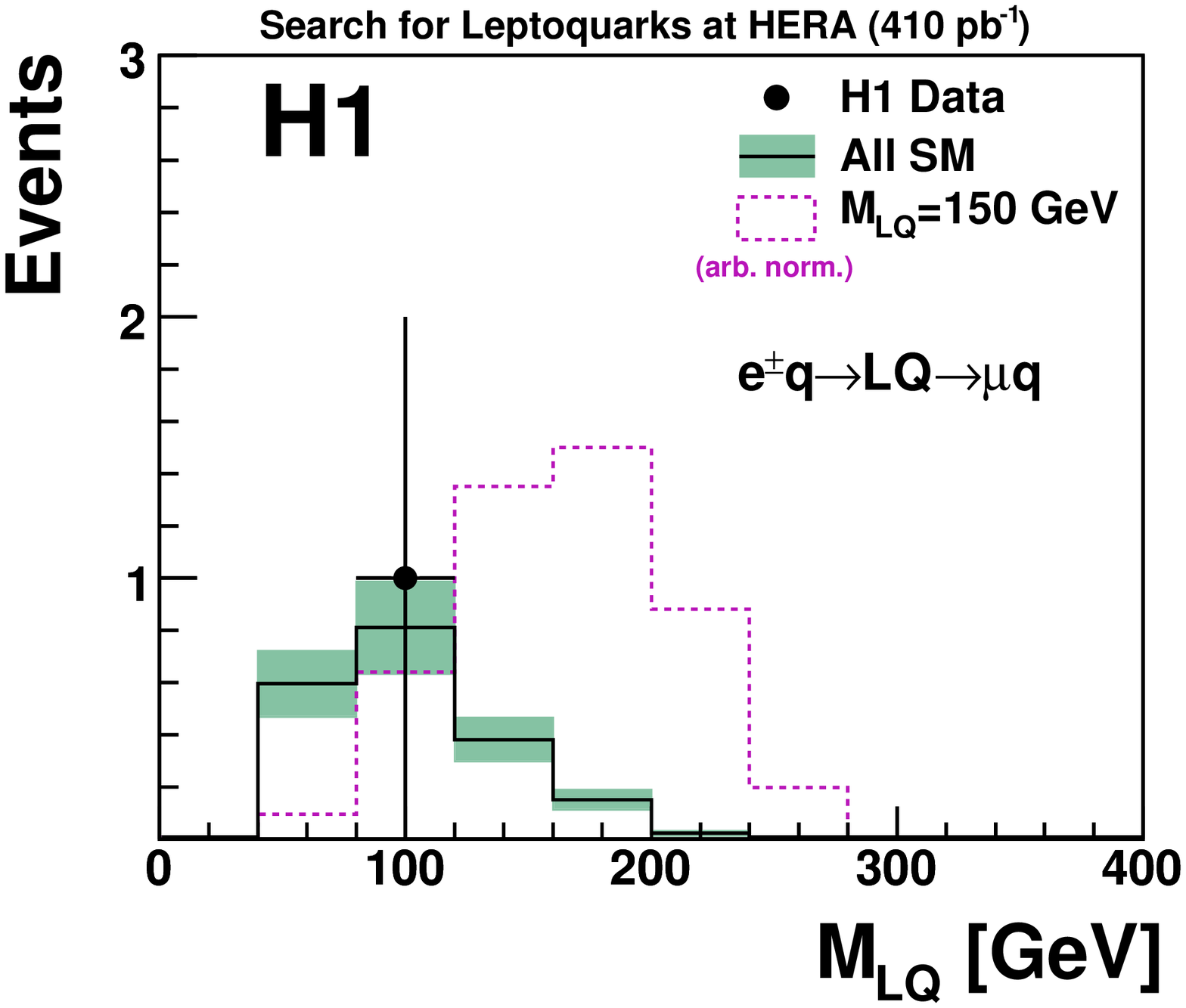}     
  \includegraphics[width=0.495\textwidth]{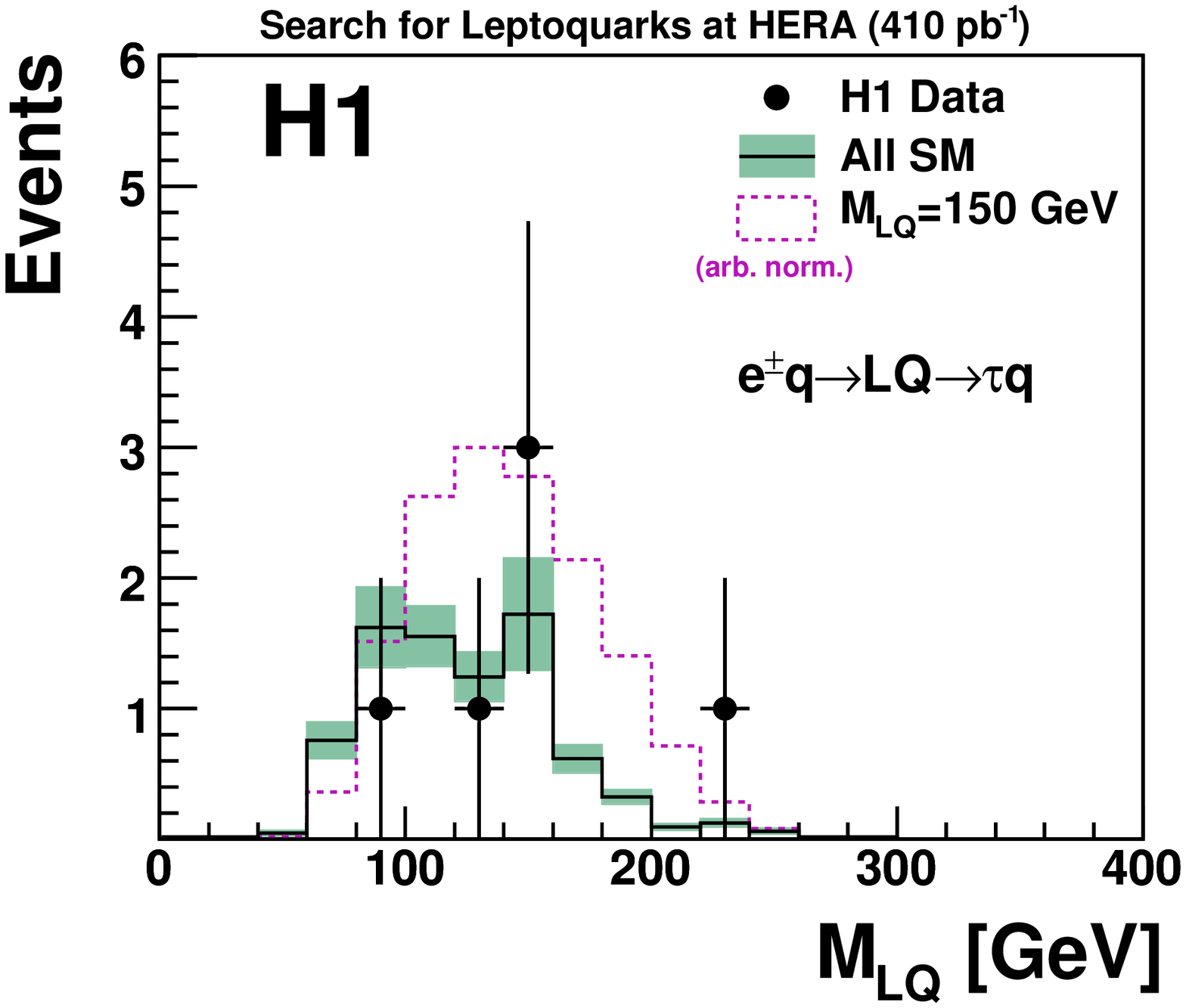}     
  \begin{picture} (0.,0.)
    \setlength{\unitlength}{1.0cm}
    \put (-9.5,2){\bf\normalsize (a)}
    \put (-2,2){\bf\normalsize (b)}
  \end{picture}
  \caption{The reconstructed leptoquark mass in the search for
    $ep \rightarrow \mu X$~(a) and $ep \rightarrow \tau X$~(b)
    events. The data are the points and the total uncertainty on the
    SM expectation (open histogram) is given by the shaded band. The
    dashed histogram indicates the LQ signal with arbitrary
    normalisation for a leptoquark mass of $150$~GeV.}
  \label{fig:massplots}
\end{figure}

After all selection cuts, the observed number of events is in agreement
with the SM prediction and therefore no evidence for LFV is found.
The reconstructed leptoquark candidate mass in the search for $ep \rightarrow \mu X$ and
$ep \rightarrow \tau X$ events is shown in figure~\ref{fig:massplots}, compared to
the SM prediction and an example LQ signal with arbitrary normalisation. 

%%%%%%%%%%%%%%%%%%%%%%%%%%%%%%%%%%%%%%%%%%%%%%%%%%%%%%%%%%%%
\section{Second and third generation leptoquark limits}

In the absence of a signal, the results of the search are interpreted in terms of
exclusion limits on the mass and the coupling of LQs mediating LFV using
a modified frequentist method with a likelihood ratio as the test statistic. 
The LQ production mechanism at HERA involves a non-zero coupling to the
first generation fermions \mbox{$\lambda_{eq_{i}} > 0$}.
For the LFV leptoquark decay, it is assumed that only one of the couplings
$\lambda_{\mu q_{j}}$ and $\lambda_{\tau q_{j}}$ is non-zero and that
$\lambda_{eq_{i}} = \lambda_{\mu q_{j}}(\lambda_{\tau q_{j}})$.

%%%

\begin{figure}[]
  \includegraphics[width=0.495\textwidth]{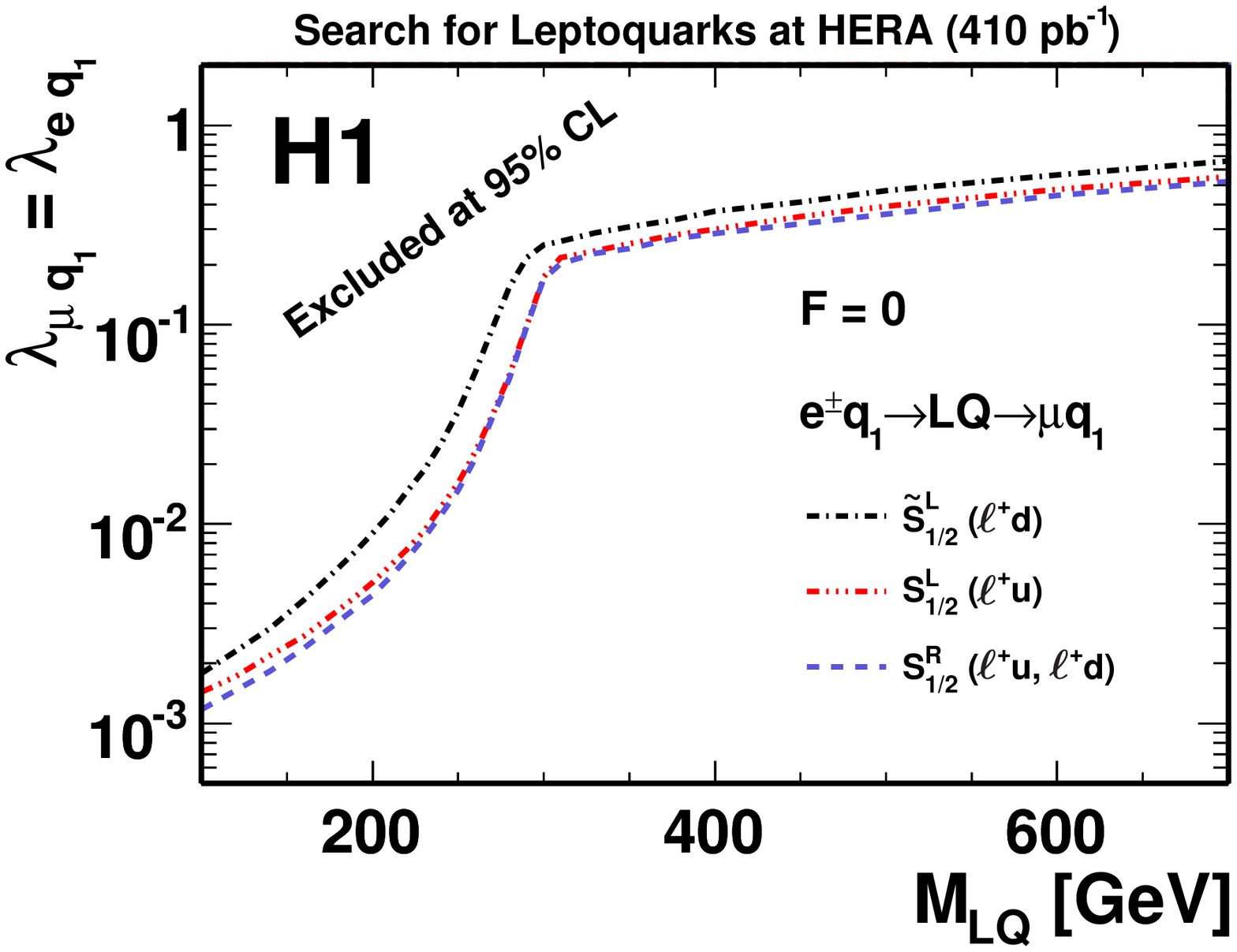}     
  \includegraphics[width=0.495\textwidth]{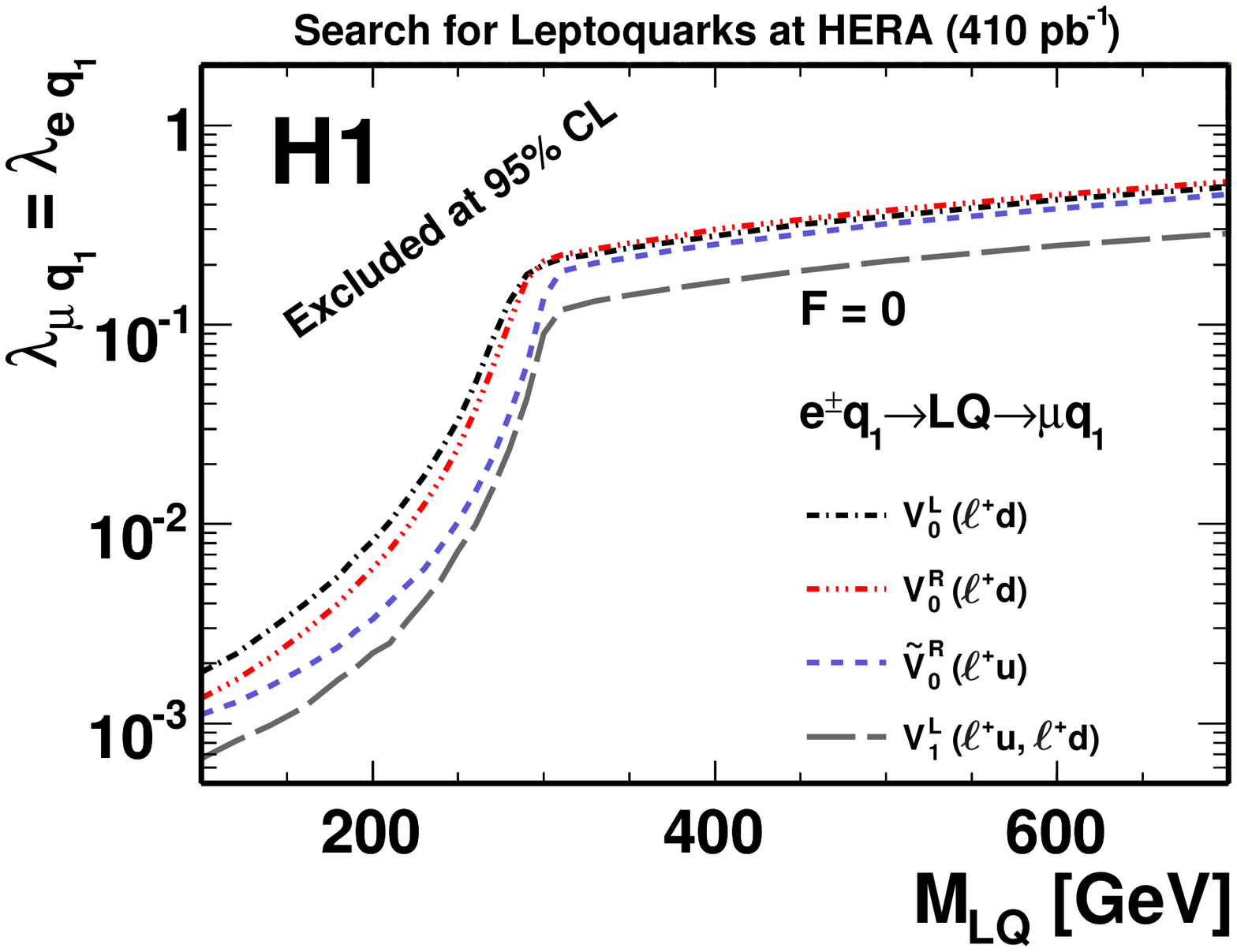}
  \includegraphics[width=0.495\textwidth]{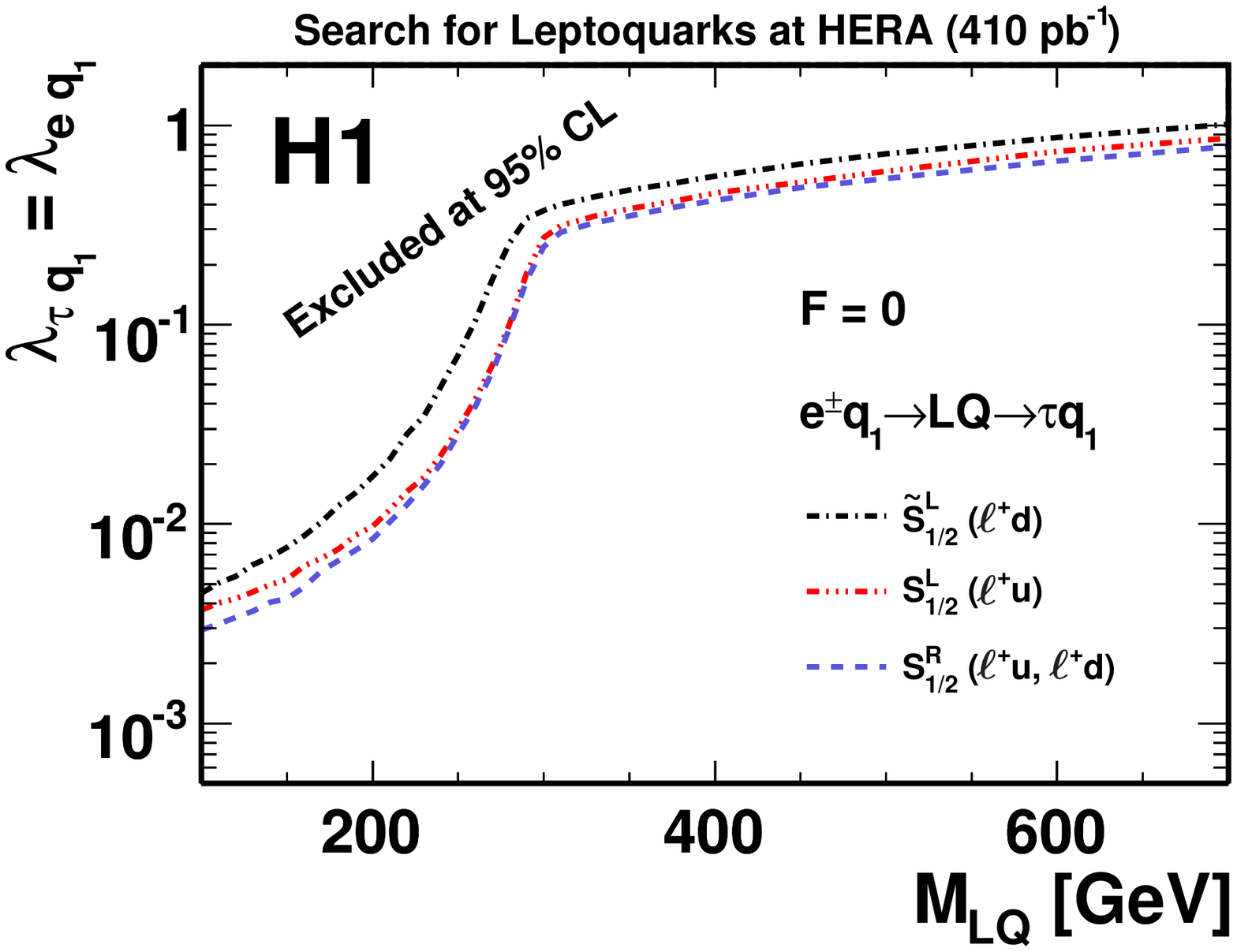}     
  \includegraphics[width=0.495\textwidth]{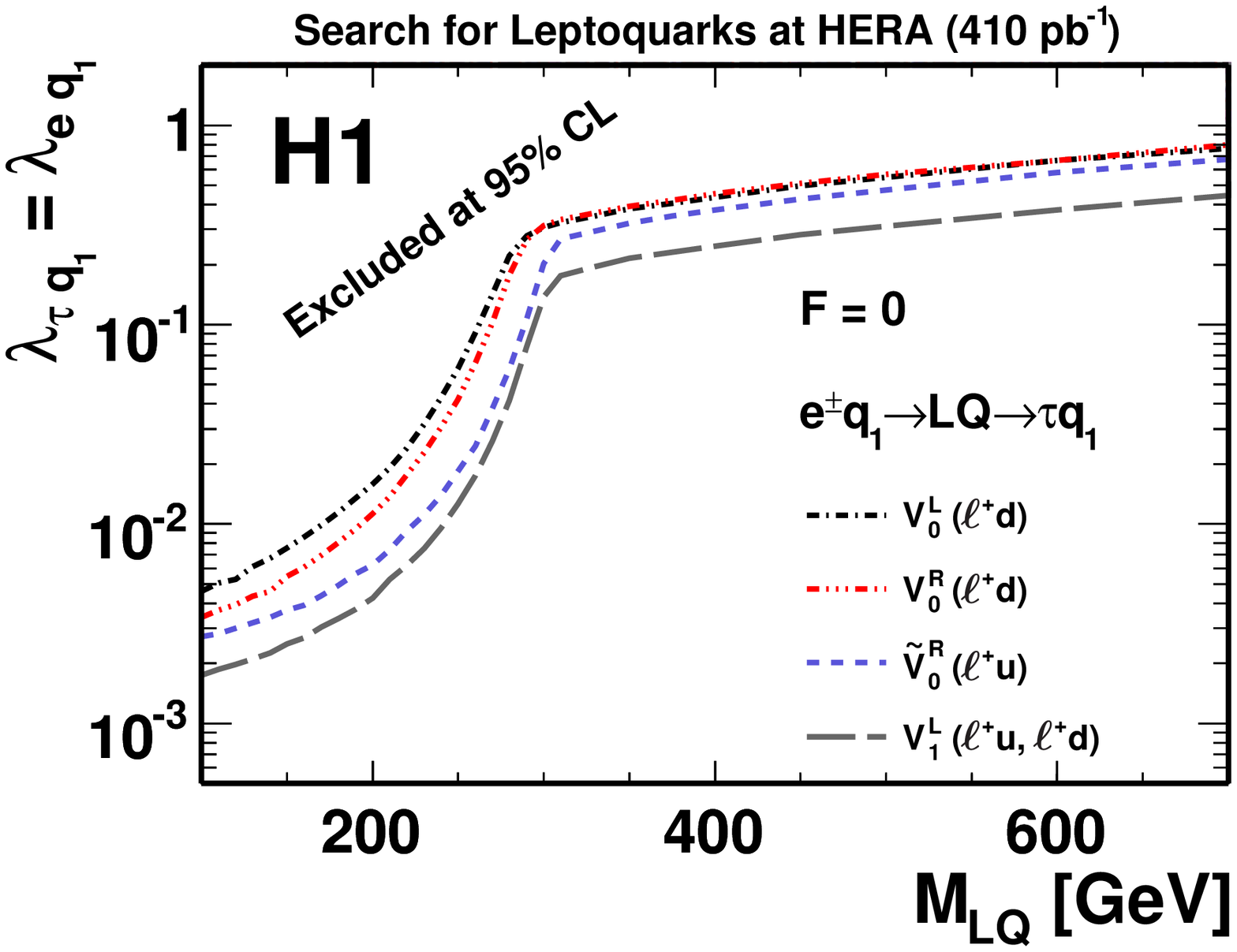}     
  \begin{picture} (0.,0.)
    \setlength{\unitlength}{1.0cm}
    \put (3.4,7.8){\bf\normalsize (a)}
    \put (11,7.8){\bf\normalsize (b)}
    \put (3.4,2.4){\bf\normalsize (c)}
    \put (11,2.4){\bf\normalsize (d)}
  \end{picture}
  \vspace{-0.4cm}
  \caption{Exclusion limits on the coupling constants $\lambda_{\ell q_{1}} = \lambda_{eq_{1}}$
    as a function of the leptoquark mass M$_{\rm LQ}$ for $F = 0$ leptoquarks:
    limits on second generation ($\ell = \mu$) scalar (a) vector (b) LQs;
    limits on third generation ($\ell = \tau$) scalar (c) vector (d) LQs.
    Regions above the lines are excluded at $95\%$~CL. The notation $q_1$ indicates that only
    processes involving first generation quarks are considered. The parentheses after the LQ name
    indicate the fermion pairs coupling to the LQ, where pairs involving anti-quarks are not shown.}
  \label{fig:limits}
\end{figure}

Figure~\ref{fig:limits} shows the $95\%$ CL upper limits on the
couplings $\lambda_{\mu q_{1}}$ and $\lambda_{\tau q_{1}}$ for $F = 0$ LQs as a
function of the mass of the LQ leading to LFV in $ep$ collisions.
Similar limits are found for $F = 2$ LQs~\cite{h1lfv2011}.
Only first generation quarks are considered here, limits involving other
quark flavours can be found in the H1 publication~\cite{h1lfv2011}.
Limits corresponding to LQs coupling to a $u$ quark are more stringent than
those corresponding  to LQs coupling to the $d$ quark only, as expected from the
larger $u$ quark density in the proton.
Corresponding to the steeply falling parton density function for high values of $x$,
the LQ production cross section decreases rapidly and exclusion limits are less stringent
towards higher LQ masses.
For LQ masses near the kinematic limit of $319$~GeV, the limit corresponding
to a resonantly  produced LQ turns smoothly into a limit on the virtual effects of both an
off-shell $s$-channel LQ process and a $u$-channel LQ exchange.
For LQ masses much greater than the HERA centre-of-mass energy the two processes
contract to an effective four-fermion interaction.

%%%

For $\lambda = \sqrt{4 \pi \alpha_{\rm em}} = 0.3$, LFV leptoquarks produced in $ep$
collisions decaying to a muon-quark or a tau-quark pair are excluded at
$95\%$~CL up to leptoquark masses of $712$~GeV and
$479$~GeV, respectively.
The H1 limits at large couplings extend beyond those reported in searches for
LQ pair production by the ATLAS~\cite{ATLAS:2012aq} and
CMS~\cite{CMS:2012dn,Chatrchyan:2012sv} Collaborations.
For $\beta_{\ell}=0.5$ LQs, which is the most appropriate value to take when comparing
LHC results to those from H1, a search for second generation scalar LQs by
CMS~\cite{CMS:2012dn} rules out LQs below $650$~GeV, at which mass this analysis
rules out scalar LQs with couplings in the range $\lambda = 0.5$-$1.0$.
The H1 limits also remain competitive with indirect limits from low-energy
experiments.

%%%%%%%%%%%%%%%%%%%%%%%%%%%%%%%%%%%%%%%%%%%%%%%%%%%%%%%%%%%%

\end{document}